\documentclass[conference]{IEEEtran}
\IEEEoverridecommandlockouts
\usepackage{cite}
\usepackage{amsmath,amssymb,amsfonts}
\usepackage{amsthm}
\usepackage{algorithmic}
\usepackage{graphicx}
\usepackage{textcomp}
\usepackage{xcolor}
\usepackage{float}
\usepackage{stfloats}
\usepackage{url}
\usepackage{subcaption}
\usepackage{soul}

\usepackage{tikz}
\usetikzlibrary{arrows.meta}

\newtheorem{definition}{Definition}

\def\BibTeX{{\rm B\kern-.05em{\sc i\kern-.025em b}\kern-.08em
    T\kern-.1667em\lower.7ex\hbox{E}\kern-.125emX}}

\begin{document}

\title{Generative Markov Modeling Framework \\ for Distributed Computing Systems
}

\author{\IEEEauthorblockN{Alfreds Lapkovskis, Ali Beikmohammadi, Sindri Magnússon, and Praveen Kumar Donta}
\IEEEauthorblockA{\textit{Department of Computer Systems and Sciences}, \textit{Stockholm University}, Stockholm 164 25, Sweden}
\texttt{\{alfreds.lapkovskis, beikmohammadi, sindri.magnusson, praveen\}@dsv.su.se}

}

\maketitle

\begin{abstract}
Emerging distributed computing paradigms, such as the computing continuum, are inherently heterogeneous, stochastic, and complex. Efficiently and effectively utilizing all available resources across the continuum demands a unified formal model of the system. To address this gap, we propose a general framework for modeling distributed computing systems as a generative Markov model, factorized over a structured system state. In our model, the state decomposes into high-dimensional variables, each further factorized over its elements, reflecting the sparse dependency structure inherent to distributed systems. This yields a tractable model that enables simulation, inference, and policy learning over otherwise intractable system states, bridging distributed computing, Markov chain theory, and reinforcement learning (RL). We demonstrate our framework through a case study of collaborative artificial intelligence (AI) inference, in which a dedicated server combines resources with those volunteered by service users. Our results show that centralized scheduling becomes a bottleneck at scale, while distributing computation across user devices reduces both latency and server resource consumption. These findings demonstrate the framework's utility for modeling, simulation, and optimization and highlight the value of adaptive decision-making in distributed computing systems.
\end{abstract}

\begin{IEEEkeywords}
Distributed computing systems, computing continuum, distributed inference, Markov model
\end{IEEEkeywords}

\section{Introduction}

The cloud computing paradigm dominates artificial intelligence (AI) inference due to advantages such as on-demand provisioning, rapid elasticity, and pay-as-you-go pricing~\cite{mell2011nist}. However, serving growing inference demand requires proportionally more infrastructure, which becomes costly at scale and carries a substantial environmental burden~\cite{li2025making}. In the age of AI, it is therefore critical to investigate more efficient alternatives.

One such alternative, actively explored in the past, is volunteer computing, a paradigm that harnesses idle resources on users' devices toward a shared computational goal~\cite{mengistu2019survey,ma2024research}. Despite renewed interest~\cite{ma2024research}, systems relying on volunteer contributions still struggle to provide stable quality-of-service (QoS) guarantees due to resource heterogeneity and intermittent user availability~\cite{mengistu2019survey}. More recently, the focus has shifted toward edge computing and computing continuum, paradigms that integrate diverse resources between user devices and cloud data centers~\cite{pujol2023edge}.
However, computing continuum systems are complex, and their behavior significantly depends on the underlying dynamic infrastructure. This makes traditional methodologies inadequate for such systems~\cite{dustdar2023distributed}. Consequently, large-scale deployments remain scarce due to both technical and economic barriers~\cite{mohan2020pruning}.
Efficiently and effectively utilizing all available resources across such systems demands a unified formal model that accounts for resource heterogeneity, stochastic availability, and the high dimensionality of system state. Without this model, reasoning about performance trade-offs and optimizing decision-making policies at scale remains difficult.

To address this gap, we propose modeling distributed computing systems as a generative Markov model over a structured system state.
The state decomposes into typed, multidimensional variables representing heterogeneous phenomena, such as workloads, resources, and execution states. Their dimensions index entities to which that phenomenon applies, such as nodes, tasks, subtasks, and resource types, while their elements represent entity-specific instances. Probabilistic Markov transitions capture stochastic system evolution. To exploit distributed locality, the transition model first factorizes over variables and then over their elements, with each factor depending only on the relevant state components. This two-level sparse factorization makes otherwise intractable states computationally manageable, enabling simulation, inference, policy evaluation, and learning. While general, the framework is primarily motivated by computing continuum systems, whose behavior is complex and infrastructure-dependent.

Prior work formulates specific computing-continuum decision problems as Markov decision processes (MDPs)~\cite{flegkas2025multiobjective,pour2025tiny}. However, as system complexity grows, narrowly scoped models may omit relevant variables and interactions, producing biased representations of the underlying dynamics. Problem-specific formulations are also difficult to reuse for other decisions. Furthermore, without a sufficiently complete model, learning effective policies requires an infeasible amount of system interaction, especially if the system dynamics evolve. Our framework instead provides a reusable system-level model that captures the shared dynamics underlying diverse distributed decision problems. Factored MDPs and dynamic Bayesian networks provide related general-purpose formalisms~\cite{guestrin2003efficient,murphy2002dbns}; however, they do not prescribe distributed computing system representations nor make their application to this domain self-evident. Our framework provides a domain-specific abstraction and structure designed around the heterogeneity, stochasticity, dynamism, and locality of distributed systems.

To demonstrate the framework's applicability, we create a concrete instantiation for a collaborative AI inference system that combines dedicated server resources with volunteered resources from service users. Under low demand, the server handles inference requests directly, ensuring high QoS. As load grows, computation is progressively offloaded to users within their resource constraints, enabling automatic scaling without proportional growth in infrastructure costs.

Our key contributions can be summarized as follows:

\begin{itemize}
    \item We propose a general generative Markov modeling framework for distributed computing systems, in which heterogeneous, high-dimensional system states are represented via a two-level sparse factorization over variables and their elements, with elements indexed by relevant system entities such as nodes, tasks, subtasks, and resource types. This structure enables tractable simulation, inference, policy evaluation, and learning.

    \item We demonstrate how the general framework can be systematically instantiated by developing a comprehensive model of a collaborative distributed AI inference system. The instantiation captures interacting system components, including stochastic user availability, dynamic workloads and resources, execution progress, resource consumption, and decision-making policies.

    \item We demonstrate the practical utility of the instantiated model through simulation-based policy evaluation. By comparing centralized and distributed scheduling under varying system configurations, we show that the model can reveal relevant performance trade-offs and support the formulation of scheduler-policy and dedicated-resource optimization problems.
\end{itemize}

The remainder of this paper is organized as follows. Section~\ref{sec:framework} introduces the proposed modeling framework. Section~\ref{sec:inference} presents its instantiation for the collaborative distributed AI inference case study. Section~\ref{sec:evaluation} presents the evaluation, and Section~\ref{sec:discussion} discusses the findings and limitations. Section~\ref{sec:conclusion} concludes the paper.

\section{Proposed Modeling Framework} \label{sec:framework}

We first establish a general framework for modeling distributed computing systems, which Section~\ref{sec:inference} then instantiates for a collaborative inference use case. In the following, we describe the proposed modeling framework. To provide better intuition, Fig.~\ref{fig:model_overview} visualizes the three levels of the framework as graphical models~\cite{michael2004graphical}.

The state of a distributed computing system at time $t$ is a tuple of $N \in
\mathbb{N}$ variables
\begin{equation}
    \mathbf{s}_t \triangleq \langle \mathbf{v}_t^{(1)}, \dots, \mathbf{v}_t^{(N)} \rangle,
    \label{eq:general_system_state}
\end{equation}
where each $\mathbf{v}_t^{(n)}$ is a high-dimensional variable with scalar or multi-dimensional elements $v_t \in
\mathbf{v}_t^{(n)}$, taking values in a domain specific to that variable. To
capture the temporal evolution of the system, we model the sequence of states
$(\mathbf{s}_t)_{t=0}^T$ as a joint distribution, which we refer to as a
\emph{generative model}. To make this tractable, we impose the Markov property.
This yields the following definition.

\begin{definition}[Generative Markov Model]\label{def:gmm}
    A generative Markov model $\mathcal{M}_T$ of a distributed computing system
    is a joint distribution over a state sequence $(\mathbf{s}_t)_{t=0}^T$ at
    discrete time steps, given by
    \begin{equation}
        \mathcal{M}_T(\mathbf{s}_0,\dots,\mathbf{s}_T) \triangleq
        p(\mathbf{s}_0)\prod_{t=1}^T p(\mathbf{s}_t \mid \mathbf{s}_{t-1}).
    \end{equation}
\end{definition}

This form imposes the Markov property, limiting dependence to the immediate previous state, and enables sequential sampling via $\mathbf{s}_t \sim p(\cdot \mid \mathbf{s}_{t-1})$. Fig.~\ref{fig:top_level} visualizes this dependency structure as a graphical model.

We assume that the model evolves at regular discrete time intervals, as observed from the perspective of the node on which it is executed. Consequently, observations collected at irregular intervals must be aligned with these intervals, for example, through resampling and, where appropriate, interpolation, before being used to estimate the model factors. During real-time operation, future states are similarly predicted at successive regular intervals from the current time step. The interval length is application-dependent and determines the model's temporal resolution.

Treating all system variables as mutually dependent within $p(\mathbf{s}_t \mid \mathbf{s}_{t-1})$ is computationally infeasible. The structure of a distributed system, however, ensures that each component evolves based on a small subset of other components rather than on the global state, which we capture as follows.

\begin{definition}[Variable Factor]
    The variable factor of $\mathbf{v}_t \in \mathbf{s}_t$ is a conditional
    distribution
    \begin{equation}
        f(\mathbf{v}_t) \triangleq p(\mathbf{v}_t \mid \mathrm{Pa}(\mathbf{v}_t)),
        \quad \mathrm{Pa}(\mathbf{v}_t) \subseteq \mathbf{s}_t \cup \mathbf{s}_{t-1},
    \end{equation}
    where $\mathbf{s}_{-1}\triangleq\emptyset$, there exists a topological ordering $\prec$ on $\mathbf{s}_t$ such that for all $\mathbf{v}, \mathbf{v}' \in \mathbf{s}_t$, $\mathbf{v} \in \mathrm{Pa}(\mathbf{v}')$ implies $\mathbf{v} \prec \mathbf{v}'$, and $p(\mathbf{s}_t \mid \mathbf{s}_{t-1}) = \prod_{\mathbf{v}_t \in \mathbf{s}_t} f(\mathbf{v}_t)$.
\end{definition}

Informally, each variable may depend on variables from the previous time step and on variables that precede it within the current time step, thereby excluding circular dependencies. Fig.~\ref{fig:second_level} illustrates this factorization of the dependency structure.

Furthermore, distributed system variables such as node states or resource
allocations are naturally indexed over nodes, tasks, and time, making their
elements sparsely dependent.
We therefore factor them at the finest granularity permitted by the dependency structure.

\begin{definition}[Element Factor]
    The element factor of $v_t \in \mathbf{v}_t$ is a conditional distribution
    \begin{equation}
        g(v_t) \triangleq p(v_t \mid \mathrm{Pa}(v_t)),
        \quad \mathrm{Pa}(v_t) \subseteq \{v \mid v \in \mathbf{v},\,
        \mathbf{v} \in \mathrm{Pa}(\mathbf{v}_t)\},
    \end{equation}
    such that $f(\mathbf{v}_t) = \prod_{v_t \in \mathbf{v}_t} g(v_t)$.
\end{definition}

Informally, each element of a high-dimensional variable depends only on a subset of the elements of its parent variables. Fig.~\ref{fig:element_factorization_1} illustrates such dependencies between elements of the same variable across consecutive time steps, whereas Fig.~\ref{fig:element_factorization_2} illustrates dependencies between elements of different variables within the same time step.

\begin{figure*}[!t]
    \centering
    \begin{subfigure}[b]{\textwidth}
        \centering
        \resizebox{\linewidth}{!}{%
        \begin{tikzpicture}[
            state/.style={circle, draw, minimum size=1.5cm, inner sep=0pt, font=\Large},
            ->, >=Stealth, shorten >=1pt, shorten <=1pt, thick
        ]
        \node[state] (s1) at (0, 0)   {$\mathbf{s}_{t-2}$};
        \node[state] (s2) at (3.025, 0)   {$\mathbf{s}_{t-1}$};
        \node[state] (s3) at (6.05, 0)   {$\mathbf{s}_t$};
        \node[state] (s4) at (9.075, 0)   {$\mathbf{s}_{t+1}$};
        \node[state] (s5) at (12.1, 0)  {$\mathbf{s}_{t+2}$};
        \draw (s1) -- (s2);
        \draw (s2) -- (s3);
        \draw (s3) -- (s4);
        \draw (s4) -- (s5);
        \draw[dotted] (-3.025, 0) -- (s1);
        \draw[dotted] (s5) -- (15.125, 0);
        \end{tikzpicture}%
        }
        \caption{
        On the top level, the generative Markov model from Definition~\ref{def:gmm} represents a joint probability distribution over a sequence of system states $\mathbf{s}_0,\dots,\mathbf{s}_t,\dots$, where at each time step $t$ the next state $\mathbf{s}_{t+1}$ is independent of previous states $\mathbf{s}_0,\dots,\mathbf{s}_{t-1}$ given the current state $\mathbf{s}_t$ (Markov property), i.e., knowing current state is sufficient to predict future states and information about the past states does not improve the prediction. This graphical model visualizes this property. The arrows $\mathbf{s}_t \rightarrow \mathbf{s}_{t+1}$ should be read as $\mathbf{s}_{t+1}$ is conditionally dependent on $\mathbf{s}_{t}$.
        }
        \label{fig:top_level}
    \end{subfigure}
 
    \vspace{6pt}

    \begin{subfigure}[b]{\textwidth}
        \centering
        \resizebox{\linewidth}{!}{%
        \begin{tikzpicture}[
            state/.style={circle, draw, minimum size=1.3cm, inner sep=0pt, font=\large},
            ->, >=Stealth, thick
        ]
 
        \node[state] (v1a) at (0, 2.4)  {$\mathbf{v}^{(1)}_{t-1}$};
        \node[state] (v2a) at (0, 0.8)  {$\mathbf{v}^{(2)}_{t-1}$};
        \node[state] (v3a) at (0,-0.8)  {$\mathbf{v}^{(3)}_{t-1}$};
        \node[state] (v4a) at (0,-2.4)  {$\mathbf{v}^{(4)}_{t-1}$};
 
        \draw (v1a) -- (v2a);
        \draw (v3a) -- (v2a);
        \draw (v3a) -- (v4a);
        \draw (v1a) to[bend left=45] (v3a);
        \draw (v2a) to[bend left=45] (v4a);
 
        \draw[dashed, rounded corners] (-1.6,-3.2) rectangle (1.1,3.2);
        \node[font=\large] at (0.75, 3.55) {$\mathbf{s}_{t-1}$};
 
        \node[state] (v1b) at (5, 2.4)  {$\mathbf{v}^{(1)}_{t}$};
        \node[state] (v2b) at (5, 0.8)  {$\mathbf{v}^{(2)}_{t}$};
        \node[state] (v3b) at (5,-0.8)  {$\mathbf{v}^{(3)}_{t}$};
        \node[state] (v4b) at (5,-2.4)  {$\mathbf{v}^{(4)}_{t}$};
 
        \draw (v1b) -- (v2b);
        \draw (v3b) -- (v2b);
        \draw (v3b) -- (v4b);
        \draw (v1b) to[bend left=45] (v3b);
        \draw (v2b) to[bend left=45] (v4b);
 
        \draw[dashed, rounded corners] (3.4,-3.2) rectangle (6.1,3.2);
        \node[font=\large] at (5.75, 3.55) {$\mathbf{s}_{t}$};
 
        \node[state] (v1c) at (10, 2.4)  {$\mathbf{v}^{(1)}_{t+1}$};
        \node[state] (v2c) at (10, 0.8)  {$\mathbf{v}^{(2)}_{t+1}$};
        \node[state] (v3c) at (10,-0.8)  {$\mathbf{v}^{(3)}_{t+1}$};
        \node[state] (v4c) at (10,-2.4)  {$\mathbf{v}^{(4)}_{t+1}$};
 
        \draw (v1c) -- (v2c);
        \draw (v3c) -- (v2c);
        \draw (v3c) -- (v4c);
        \draw (v1c) to[bend left=45] (v3c);
        \draw (v2c) to[bend left=45] (v4c);
 
        \draw[dashed, rounded corners] (8.4,-3.2) rectangle (11.1,3.2);
        \node[font=\large] at (10.75, 3.55) {$\mathbf{s}_{t+1}$};
 
        \draw (v1a) -- (v1b);
        \draw (v2a) -- (v2b);
        \draw (v3a) -- (v3b);
        \draw (v4a) -- (v4b);
        \draw (v2a) -- (v1b);
        \draw (v4a) -- (v3b);
 
        \draw (v1b) -- (v1c);
        \draw (v2b) -- (v2c);
        \draw (v3b) -- (v3c);
        \draw (v4b) -- (v4c);
        \draw (v2b) -- (v1c);
        \draw (v4b) -- (v3c);
 
        \draw[dotted] (-3.0, 2.4)  -- (v1a);
        \draw[dotted] (-3.0, 0.8)  -- (v2a);
        \draw[dotted] (-3.0,-0.8)  -- (v3a);
        \draw[dotted] (-3.0,-2.4)  -- (v4a);
        \draw[dotted] (-3.0, 1.44) -- (v1a);
        \draw[dotted] (-3.0,-1.76) -- (v3a);
 
        \draw[dotted] (v1c) -- (13.0, 2.4);
        \draw[dotted] (v2c) -- (13.0, 0.8);
        \draw[dotted] (v3c) -- (13.0,-0.8);
        \draw[dotted] (v4c) -- (13.0,-2.4);
        \draw[dotted] (v2c) -- (13.0, 1.76);
        \draw[dotted] (v4c) -- (13.0,-1.44);
 
        \end{tikzpicture}%
        }
        \caption{
        On the second level, states $\mathbf{s}_t$ comprise (multi-dimensional) system variables $\mathbf{v}^{(1)}_t,\dots,\mathbf{v}^{(N)}_t$, where each variable corresponds to a certain system characteristic. Variables can represent availability of system components or resources, task states, actions, etc. Essentially, variables may encode anything characterized by dynamically changing values. The dependencies between variables are generally sparse; i.e., a variable should not depend on all other variables. Moreover, a variable may depend only on other variables from the current or previous time step. 
        This graphical model exemplifies factorization of Markovian state dependencies into sparse dependencies among individual variables.
        }
        \label{fig:second_level}
    \end{subfigure}
 
    \vspace{6pt}

    \begin{subfigure}[b]{0.48\textwidth}
        \centering
        \resizebox{\linewidth}{!}{%
        \begin{tikzpicture}[
            state/.style={circle, draw, minimum size=0.8cm, inner sep=0pt},
            ->, >=Stealth, thick
        ]
 
        \node[state] (A1a) at (0, 3.20)  {\tiny $j=1$};
        \node[state] (A2a) at (0, 2.20)  {\tiny $j=2$};
        \node[state] (B1a) at (0, 0.5)  {\tiny $j=1$};
        \node[state] (B2a) at (0,-0.5)  {\tiny $j=2$};
        \node[state] (C1a) at (0,-2.20)  {\tiny $j=1$};
        \node[state] (C2a) at (0,-3.20)  {\tiny $j=2$};
        \draw[dashed, rounded corners] (-1.0,-4.20) rectangle (1.0,4.20);
        \node[font=\normalsize] at (0, 4.50) {$\mathbf{v}^{(1)}_{t-1}$};
 
        \node[state] (A1b) at (4.5, 3.20)  {\tiny $j=1$};
        \node[state] (A2b) at (4.5, 2.20)  {\tiny $j=2$};
        \node[state] (B1b) at (4.5, 0.5)  {\tiny $j=1$};
        \node[state] (B2b) at (4.5,-0.5)  {\tiny $j=2$};
        \node[state] (C1b) at (4.5,-2.20)  {\tiny $j=1$};
        \node[state] (C2b) at (4.5,-3.20)  {\tiny $j=2$};
        \draw[dashed, rounded corners] (3.5,-4.20) rectangle (5.5,4.20);
        \node[font=\normalsize] at (4.5, 4.50) {$\mathbf{v}^{(1)}_{t}$};
 
        \node[state] (A1c) at (9, 3.20)  {\tiny $j=1$};
        \node[state] (A2c) at (9, 2.20)  {\tiny $j=2$};
        \node[state] (B1c) at (9, 0.5)  {\tiny $j=1$};
        \node[state] (B2c) at (9,-0.5)  {\tiny $j=2$};
        \node[state] (C1c) at (9,-2.20)  {\tiny $j=1$};
        \node[state] (C2c) at (9,-3.20)  {\tiny $j=2$};
        \draw[dashed, rounded corners] (8.0,-4.20) rectangle (10.0,4.20);
        \node[font=\normalsize] at (9, 4.50) {$\mathbf{v}^{(1)}_{t+1}$};
 
        \foreach \x in {0, 4.5, 9} {
          \foreach \ymid in {2.7, 0.0, -2.7} {
            \pgfmathsetmacro{\ytop}{\ymid+1.2}
            \pgfmathsetmacro{\ybot}{\ymid-1.2}
            \pgfmathsetmacro{\xl}{\x-0.7}
            \pgfmathsetmacro{\xr}{\x+0.7}
            \draw[dotted, rounded corners] (\xl,\ybot) rectangle (\xr,\ytop);
          }
        }
        \foreach \x in {0, 4.5, 9} {
          \node[font=\scriptsize] at (\x, 4.05) {$i=1$};
          \node[font=\scriptsize] at (\x, 1.35) {$i=2$};
          \node[font=\scriptsize] at (\x,-1.35) {$i=3$};
        }
 
        \foreach \X in {A,B,C} {
          \draw (\X1a) -- (\X1b);
          \draw (\X1a) -- (\X2b);
          \draw (\X2a) -- (\X1b);
          \draw (\X2a) -- (\X2b);
        }
        \foreach \X in {A,B,C} {
          \draw (\X1b) -- (\X1c);
          \draw (\X1b) -- (\X2c);
          \draw (\X2b) -- (\X1c);
          \draw (\X2b) -- (\X2c);
        }
 
        \coordinate (vA1) at (-2.2, 3.20);
        \coordinate (vA2) at (-2.2, 2.20);
        \coordinate (vB1) at (-2.2, 0.5);
        \coordinate (vB2) at (-2.2,-0.5);
        \coordinate (vC1) at (-2.2,-2.20);
        \coordinate (vC2) at (-2.2,-3.20);
        \foreach \X in {A,B,C} {
          \draw[dotted] (v\X1) -- (\X1a);
          \draw[dotted] (v\X1) -- (\X2a);
          \draw[dotted] (v\X2) -- (\X1a);
          \draw[dotted] (v\X2) -- (\X2a);
        }
 
        \node[circle, minimum size=0.18cm, inner sep=0pt] (wA1) at (11.2, 3.20) {};
        \node[circle, minimum size=0.18cm, inner sep=0pt] (wA2) at (11.2, 2.20) {};
        \node[circle, minimum size=0.18cm, inner sep=0pt] (wB1) at (11.2, 0.5) {};
        \node[circle, minimum size=0.18cm, inner sep=0pt] (wB2) at (11.2,-0.5) {};
        \node[circle, minimum size=0.18cm, inner sep=0pt] (wC1) at (11.2,-2.20) {};
        \node[circle, minimum size=0.18cm, inner sep=0pt] (wC2) at (11.2,-3.20) {};
        \foreach \X in {A,B,C} {
          \draw[dotted] (\X1c) -- (w\X1);
          \draw[dotted] (\X1c) -- (w\X2);
          \draw[dotted] (\X2c) -- (w\X1);
          \draw[dotted] (\X2c) -- (w\X2);
        }
 
        \end{tikzpicture}%
        }
        \caption{
        On the third level, each variable $\mathbf{v}^{(k)}_t$ comprises individual elements indexed by multiple dimensions. Dimensions index some system entities to which the corresponding variable values pertain. In this example, $i$ could index nodes, and $j$ could index computational tasks. Dependencies between elements are generally sparse; e.g., a task at one node depends only on tasks at the same node.
        }
        \label{fig:element_factorization_1}
    \end{subfigure}
    \hfill
    \begin{subfigure}[b]{0.48\textwidth}
        \centering
        \resizebox{0.698\linewidth}{!}{%
        \begin{tikzpicture}[
            state/.style={circle, draw, minimum size=0.8cm, inner sep=0pt},
            ->, >=Stealth, thick
        ]
 
        \node[state] (A1) at (0, 3.20)  {\tiny $j=1$};
        \node[state] (A2) at (0, 2.20)  {\tiny $j=2$};
        \node[state] (B1) at (0, 0.5)  {\tiny $j=1$};
        \node[state] (B2) at (0,-0.5)  {\tiny $j=2$};
        \node[state] (C1) at (0,-2.20)  {\tiny $j=1$};
        \node[state] (C2) at (0,-3.20)  {\tiny $j=2$};
        \draw[dashed, rounded corners] (-1.0,-4.20) rectangle (1.0,4.20);
        \node[font=\normalsize] at (0, 4.50) {$\mathbf{v}^{(1)}_{t}$};
 
        \foreach \ymid in {2.7, 0.0, -2.7} {
          \pgfmathsetmacro{\ytop}{\ymid+1.2}
          \pgfmathsetmacro{\ybot}{\ymid-1.2}
          \draw[dotted, rounded corners] (-0.7,\ybot) rectangle (0.7,\ytop);
        }
        \node[font=\scriptsize] at (0, 4.05) {$i=1$};
        \node[font=\scriptsize] at (0, 1.35) {$i=2$};
        \node[font=\scriptsize] at (0,-1.35) {$i=3$};
 
        \node[state] (N1a) at (5, 3.2) {\tiny $m=1$};
        \node[state] (N1b) at (5, 2.2) {\tiny $m=2$};
        \node[state] (N1c) at (5, 1.2) {\tiny $m=3$};
        \node[state] (N2a) at (5,-1.2) {\tiny $m=1$};
        \node[state] (N2b) at (5,-2.2) {\tiny $m=2$};
        \node[state] (N2c) at (5,-3.2) {\tiny $m=3$};
        \draw[dashed, rounded corners] (4.0,-4.20) rectangle (6.0,4.20);
        \node[font=\normalsize] at (5, 4.50) {$\mathbf{v}^{(2)}_{t}$};
 
        \draw[dotted, rounded corners] (4.3,0.5) rectangle (5.7,3.9);
        \draw[dotted, rounded corners] (4.3,-3.9) rectangle (5.7,-0.5);
        \node[font=\scriptsize] at (5, 4.05) {$n=1$};
        \node[font=\scriptsize] at (5,-0.35) {$n=2$};
 
        \foreach \S in {A1,B1,C1} {
          \foreach \T in {N1a,N1b,N1c} {
            \draw (\S) -- (\T);
          }
        }
        \foreach \S in {A2,B2,C2} {
          \foreach \T in {N2a,N2b,N2c} {
            \draw (\S) -- (\T);
          }
        }
 
        \coordinate (vA1) at (-2.2, 3.20);
        \coordinate (vA2) at (-2.2, 2.20);
        \coordinate (vB1) at (-2.2, 0.5);
        \coordinate (vB2) at (-2.2,-0.5);
        \coordinate (vC1) at (-2.2,-2.20);
        \coordinate (vC2) at (-2.2,-3.20);
        \foreach \X in {A,B,C} {
          \draw[dotted] (v\X1) -- (\X1);
          \draw[dotted] (v\X1) -- (\X2);
          \draw[dotted] (v\X2) -- (\X1);
          \draw[dotted] (v\X2) -- (\X2);
        }
 
        \foreach \Y/\N in {3.2/N1a, 2.2/N1b, 1.2/N1c, -1.2/N2a, -2.2/N2b, -3.2/N2c} {
          \draw[dotted] (\N) -- (7.2, \Y);
        }
 
        \end{tikzpicture}%
        }
        \caption{
        Similar to temporal dependencies between elements of the same variable, as depicted in Fig.~\ref{fig:element_factorization_1}, there are sparse dependencies between elements of different variables. This graphical model illustrates this, e.g., $\mathbf{v}^{(1)}_t$ could be resource usage, where $i$ indexes resource types and $j$ nodes, and $\mathbf{v}^{(2)}_t$ could be task completion with $n$ indexing nodes and $m$ individual tasks.
        }
        
        \label{fig:element_factorization_2}
    \end{subfigure}
 
    \caption{
    Graphical models visualizing the proposed generative Markov model. (a) illustrates the top-level Markovian temporal state dependency structure. (b) illustrates the factorization of states into sparsely dependent system variables. (c) illustrates the element-wise factorization of dependencies among system variables.
    }
    \label{fig:model_overview}
\end{figure*}

\begin{figure*}%[!t]
    \centering
    \resizebox{\linewidth}{!}{%
    \begin{tikzpicture}[
        server/.style={rectangle, draw, thick, minimum width=2.2cm, minimum height=0.9cm, rounded corners=1pt},
        devOn/.style={circle, draw, thick, fill=black, minimum size=0.4cm},
        devOff/.style={circle, draw, thick, dashed, minimum size=0.4cm},
        requestArrow/.style={->, >=Stealth, thick},
        dispatchArrow/.style={->, >=Stealth, thick, densely dotted},
        resultArrow/.style={->, >=Stealth, thick, dashed},
        region/.style={draw, dashed, rounded corners=6pt},
    ]

    \node[server] (server) at (0, 5.2) {Server};

    \draw[dispatchArrow] (server.north east) .. controls +(0.5,0.5) and +(0.5,-0.5) .. (server.east);
    \node[font=\scriptsize] at (2.0, 5.45) {subtask};

    \node[devOn]  (d1)  at (-6.3, 2.25) {};
    \node[devOn]  (d2)  at (-4.9, 2.25) {};
    \node[devOff] (d3)  at (-3.5, 2.25) {};
    \node[devOn]  (d4)  at (-2.1, 2.25) {};
    \node[devOff] (d5)  at (-0.7, 2.25) {};
    \node[devOff] (d6)  at ( 0.7, 2.25) {};
    \node[devOn]  (d7)  at ( 2.1, 2.25) {};
    \node[devOn]  (d8)  at ( 3.5, 2.25) {};
    \node[devOff] (d9)  at ( 4.9, 2.25) {};
    \node[devOff] (d10) at ( 6.3, 2.25) {};

    \draw[region] (-7.0,1.75) rectangle (7.0,2.75);
    \node[font=\small] at (0, 1.3)
        {User devices ( \tikz[baseline=-0.5ex]{\draw[thick,fill=black] (0,0) circle (0.12);}\ online,
         \tikz[baseline=-0.5ex]{\draw[thick,dashed] (0,0) circle (0.12);}\ offline )};

    \draw[requestArrow, bend left=25] (d1) to (server.south west);
    \node[font=\scriptsize] at (-5.75, 3.55) {request};

    \foreach \d in {4,7,8} {
      \draw[dispatchArrow, bend right=18] (server.south) to (d\d);
    }
    \node[font=\scriptsize] at (-0.5, 3.55) {subtasks};

    \foreach \d in {7,8} {
      \draw[resultArrow, bend right=18] (d\d) to (server.south);
    }
    \node[font=\scriptsize] at (3.2, 3.55) {results};

    \end{tikzpicture}%
    }
    \caption{
    High-level overview of the collaborative distributed AI inference system considered in the case study. Online users submit inference requests to the server, whose scheduler assigns the required subtasks to dedicated server resources and idle resources contributed by online users, potentially including the requesting user. Once all subtasks of a request have been completed, the server aggregates their results and returns the final inference result to the requesting user.
    }
    \label{fig:distributed_inference_system}
\end{figure*}

Applying both levels of factorization to Definition~\ref{def:gmm}, the model takes its fully expanded form
\begin{equation}
    \mathcal{M}_T(\mathbf{s}_0,\dots,\mathbf{s}_T) \triangleq
    \prod_{0 \le t \le T}\; \prod_{\mathbf{v}_t \in \mathbf{s}_t}\;
    \prod_{v_t \in \mathbf{v}_t} g(v_t).
\end{equation}
The following section defines the concrete variables and factors for the collaborative distributed inference setting.

\section{Case Study}
\label{sec:inference}

To demonstrate the proposed modeling framework, we consider a use case of \emph{collaborative distributed AI inference}. Centralized AI inference services accommodate rising workloads by increasing server capacity. This ties service growth to the continued expansion of dedicated infrastructure. Our case study investigates whether this dependence can be reduced through a collaborative AI inference system that supplements a dedicated server with idle resources from participating users. The server provides dependable baseline capacity, while user devices form an additional dynamic resource pool whose capacity increases with user participation. 
Fig.~\ref{fig:distributed_inference_system} provides a high-level overview of such a system.
Simulation of such a system and investigation of various decision-making strategies require a model. For this purpose, we instantiate the proposed generative Markov model of Definition~\ref{def:gmm}.

Formally, the system comprises the following entities. Users $i \in \mathcal{I}$ and a dedicated server $v_0$ form the set of nodes $\mathcal{V} = \mathcal{I} \cup \{v_0\}$, each equipped with resources of types $r \in \mathcal{R}$. The system serves $|\mathcal{K}|$ inference task types, each decomposed into independent subtasks $s \in \mathcal{S}_k$. Users both submit requests of type $k \in \mathcal{K}$ and contribute idle resources to execute subtasks associated with other users' requests. Decision-making is governed by two policy factors: the \emph{scheduler} $\pi$ and the \emph{executor} $\varsigma$, which jointly produce the action $\mathbf{u}_t \triangleq \langle \mathbf{u}^{\pi}_t, \mathbf{u}^{\varsigma}_t \rangle$. We model the system in discrete time with a step size of $1\,\mathrm{s}$.

We select the state variables that capture the mechanisms governing workload, execution feasibility, progress, completion, and policy decisions. Together, these variables form a compact scheduling-relevant state
\begin{equation}
    \mathbf{s}_t = \langle \mathbf{o}_t, \mathbf{q}_t, \mathbf{a}_t, \mathbf{x}_t, \mathbf{y}_t, \mathbf{c}_t, \mathbf{d}_t, \mathbf{u}_t \rangle.
\end{equation}
The following subsection describes these variables and their child-parent relationships.

\subsection{State Variables}

\textbf{User availability} 
$\mathbf{o}_t$ represents intermittent participation, which determines whether a user can submit requests, contribute resources, or make execution progress.
It depends on the previous availability $\mathbf{o}_{t-1}$ and duration $\mathbf{d}^o_t$ measuring time since the last state change, giving $\mathrm{Pa}(\mathbf{o}_t) = \{\mathbf{o}_{t-1}, \mathbf{d}^o_t\}$ and $\mathrm{Pa}(\mathbf{d}^o_t) = \{\mathbf{d}^o_{t-1}, \mathbf{o}_{t-1}\}$.
We condition on $\mathbf{d}^o_t$ because online and offline periods need not be memoryless, i.e., the probability of leaving a state may depend on how long the user has remained in it.

We model the \textbf{request state} $\mathbf{q}_t$ separately from availability, because an online user may have no active request yet still contribute resources.
Each online user may have at most one active inference request. The request state $\mathbf{q}_t \in \mathcal{Q}^{|\mathcal{I}|}$, where $\mathcal{Q} = \mathcal{K} \cup \{k_0\}$ with $k_0$ denoting no active request, depends on $\mathbf{q}_{t-1}$ and $\mathbf{o}_t$, giving $\mathrm{Pa}(\mathbf{q}_t) = \{\mathbf{q}_{t-1}, \mathbf{o}_t, \mathbf{d}^q_t\}$. A duration variable $\mathbf{d}^q_t$ additionally depends on execution states $\mathbf{y}_{t-1}$ to detect request completion, giving $\mathrm{Pa}(\mathbf{d}^q_t) = \{\mathbf{d}^q_{t-1}, \mathbf{q}_{t-1}, \mathbf{o}_t, \mathbf{y}_{t-1}\}$.

\textbf{Allocated resources} $\mathbf{a}_t$ represent the bounded fluctuating capacity that each node contributes to the system. They are distinct from resource consumption because equal-capacity nodes may experience different loads from concurrent transfers and executions.
Offline users contribute no resources, so $\mathbf{o}_t \in \mathrm{Pa}(\mathbf{a}_t)$. Resource trajectories can be autocorrelated in practice and may not be first-order Markov on their own, so we augment the state with a history $\tilde{\mathbf{a}}_{t-1} \triangleq [\mathbf{a}_{t-1}, \dots, \mathbf{a}_{t-H}]$ of size $H$, implicitly included in $\mathbf{s}_t$, giving $\mathrm{Pa}(\mathbf{a}_t) = \{\tilde{\mathbf{a}}_{t-1}, \mathbf{o}_t\}$.

Subtask execution requires its associated data to be downloaded to the executing node first, tracked by the \textbf{readiness state} $\mathbf{x}_t$, after which subtasks progress through execution phases, tracked by the \textbf{execution state} $\mathbf{y}_t$. Both variables follow the same pattern: auxiliary variables $\hat{\mathbf{x}}_t, \hat{\mathbf{y}}_t$ (treated as deterministic intermediate quantities) apply actions $\mathbf{u}_{t-1}$ to $\mathbf{x}_{t-1},\mathbf{y}_{t-1}$, conditioned on $\mathbf{o}_t$ (since offline nodes cannot progress), giving, e.g.,  $\mathrm{Pa}(\hat{\mathbf{x}}_t) = \{\mathbf{x}_{t-1}, \mathbf{o}_t, \mathbf{u}^{\pi,x}_{t-1},\mathbf{u}^{\varsigma,x}_{t-1}\}$. Durations $\mathbf{d}^x_t,\mathbf{d}^y_t$ capture the pace of the current phase based on $\mathbf{c}_t$, giving e.g. $\mathrm{Pa}(\mathbf{d}^x_t) = \{\mathbf{d}^x_{t-1}, \hat{\mathbf{x}}_t, \mathbf{c}^x_t\}$. The actual $\mathbf{x}_t,\mathbf{y}_t$ resolve from the auxiliary $\hat{\mathbf{x}}_t,\hat{\mathbf{y}}_t$ and $\mathbf{d}^x_{t},\mathbf{d}^y_{t}$, advancing when the duration signals phase completion.

\textbf{Resource consumption} $\mathbf{c}_t \triangleq \langle \mathbf{c}^x_t, \mathbf{c}^y_t \rangle$ covers data transfers and subtask execution over $\mathcal{R}^X, \mathcal{R}^Y \subseteq \mathcal{R}$ respectively. Both components are modeled jointly per node since concurrent subtasks share resources non-additively, e.g., bandwidth is split rather than replicated. Consumption is bounded by $\mathbf{a}_t$; $\mathbf{x}_t,\mathbf{y}_t$ determine consuming subtasks, $\mathbf{d}^x_t,\mathbf{d}^y_t$ identify co-scheduled subtasks for batch execution, and history terms $\tilde{\mathbf{c}}^x_{t-1}, \tilde{\mathbf{c}}^y_{t-1}$ account for autocorrelation,
giving $\mathrm{Pa}(\mathbf{c}^x_t, \mathbf{c}^y_t) = \{\tilde{\mathbf{c}}^x_{t-1}, \tilde{\mathbf{c}}^y_{t-1}, \mathbf{a}_t, \hat{\mathbf{x}}_t, \hat{\mathbf{y}}_t, \mathbf{d}^x_{t-1}, \mathbf{d}^y_{t-1}\}$.

\subsection{Scheduler and Executor Policies}

The \textbf{scheduler} $\pi(\mathbf{u}^\pi_t \mid \mathbf{s}_t \setminus \mathbf{u}_t)$ runs on the server, deciding on actions for subtask downloads and execution assignments across nodes ($\mathbf{u}^\pi_t = \langle \mathbf{u}^{\pi,x}_t, \mathbf{u}^{\pi,y}_t \rangle$). The \textbf{executor} $\varsigma(\mathbf{u}^\varsigma_t \mid \mathbf{s}_t \setminus \mathbf{u}_t)$ runs on each node, enforcing local resource constraints via aborts and pauses ($\mathbf{u}^\varsigma_t = \langle \mathbf{u}^{\varsigma,x}_t, \mathbf{u}^{\varsigma,y}_t \rangle$). Executor actions take priority over scheduler actions to guarantee constraint compliance.

\subsection{Dimensionality and Element Factorization}

System variables are high-dimensional and are indexed over the relevant system entities, e.g., $\mathbf{x}_t \in \mathcal{X}^{|\mathcal{V}| \times |\mathcal{K}| \times S_{\max}}$ tracks readiness per node, task, and subtask ($S_{\max} \triangleq \max_k |\mathcal{S}_k|$), while $\mathbf{y}_t \in \mathcal{Y}^{|\mathcal{I}| \times |\mathcal{V}| \times |\mathcal{K}| \times S_{\max}}$ is further indexed by requesting user. Most variables, however, evolve independently across these entities, so $f(\mathbf{v}_t)$ factorizes into sparse element factors $g(v_t)$, e.g., assuming availability independence across users, $f(\mathbf{o}_t) = \prod_{i \in \mathcal{I}} p(o_{t,i} \mid o_{t-1,i}, d^o_{t,i})$.

\subsection{Problem Formulations}

Given the generative Markov model of the collaborative inference system developed earlier, we now formulate the optimization problems that govern decision-making within it.

\begin{definition}[Scheduler Policy Optimization]
Given a generative Markov model $\mathcal{M}_T$ with $T\rightarrow\infty$, a fixed executor $\varsigma$, a reward function $r(\mathbf{s}_t)$, find a scheduler policy $\pi^*$ such that
\begin{equation}
\pi^*=\arg\max_\pi\; \liminf_{T\rightarrow\infty}\,\mathbb{E}\left[\frac{1}{T} \sum_{t=1}^T r(\mathbf{s}_t) \mid \mathbf{s}_0, \mathbf{u}^\pi_{0:t-1} \sim \pi \right].
\end{equation} 
\label{def:scheduler_policy_optimization}
\end{definition}
The reward $r(\mathbf{s}_t)$ captures QoS in terms of serving latency. Specifically,
\begin{equation}
r(\mathbf{s}_t)\triangleq - \mathbf{1}^\top\boldsymbol{\delta}_t,
\end{equation}
where $\boldsymbol{\delta}_t \triangleq \boldsymbol{\delta}(\mathbf{q}_t, \mathbf{d}^q_t) \in \{0,1\}^{|\mathcal{I}|}$ are per-user latency increments. This is an average-reward RL problem~\cite{puterman2014markov,sutton2018rl}. While approachable via existing methods, its solution in this context is deferred to future work.

\begin{definition}[Dedicated Resource Optimization] \label{def:dedicated_resource_optimization}
Given fixed scheduler and executor policies, a nominal resource vector $\mathbf{a}_\dagger$ parametrizing the server resource distribution $p(a_{t,r,v_0} \mid a_{t-1,r,v_0};\, a_{\dagger,r})$ for all $r \in \mathcal{R}$, significance levels $\epsilon_k \in (0,1)$, and per-type latency thresholds $\tau_k$, find $\mathbf{a}^*_{\dagger}$ such that
\begin{equation}
    \begin{aligned}
        \mathbf{a}^*_{\dagger} = \arg\min_{\mathbf{a}_{\dagger}}&\; \lVert \mathbf{a}_{\dagger} \rVert_{\mathbf{w}} \\
        \mathrm{s.t.}& \quad c_k\!\left((\mathbf{s}_t)_{t=0}^T \sim \mathcal{M}_T\right) \ge 1-\epsilon_k, \quad \forall k\in\mathcal{K},
    \end{aligned}
\end{equation}
where $\lVert \mathbf{a}_{\dagger} \rVert_{\mathbf{w}} = \sum_{r \in \mathcal{R}} w_r\, a_{\dagger,r}$ is a weighted $\ell^1$ norm with per-resource cost weights $\mathbf{w} \in \mathbb{R}^{|\mathcal{R}|}_{>0}$, and the chance constraints are defined per request type $k$ as
\begin{equation}
    c_k\!\left((\mathbf{s}_t)_{t=0}^T\right) \triangleq \mathbb{P}\!\left(d^q_{t,i} \le \tau_k \mid q_{t+1,i} = k_0,\, q_{t,i} = k\right).
\end{equation}
\end{definition}

The nominal resource vector $\mathbf{a}_{\dagger}$ is necessary because fixing server resources in practice is not always feasible, as uncontrollable factors 
cause actual availability to deviate from any intended value. The chance constraints ensure that reducing dedicated resources does not degrade QoS below an acceptable level for any request type, with separate thresholds $\tau_k$ allowing different latency requirements per task type.

\begin{figure*}%[!ht]
    \centering
    \begin{subfigure}[t]{0.244\textwidth}
        \centering
        \includegraphics[width=\textwidth]{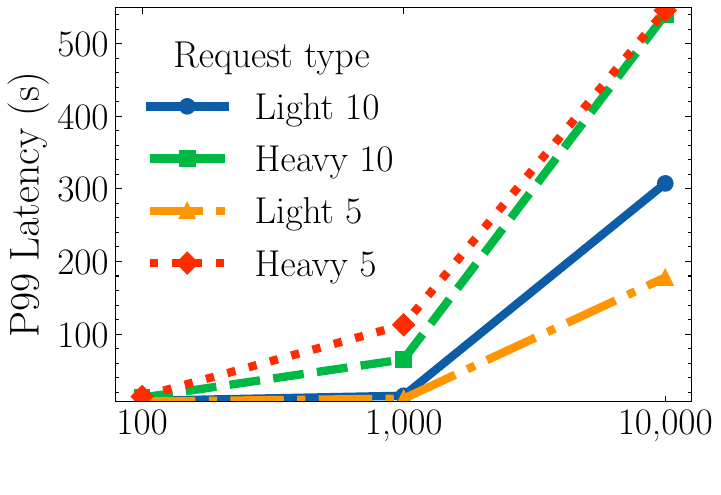}

        \vspace{-1em}
        \caption{\centering Centralized 10$\times$}
        \vspace{0.2em}
    \end{subfigure}
    \hfill
    \begin{subfigure}[t]{0.244\textwidth}
        \centering
        \includegraphics[width=\textwidth]{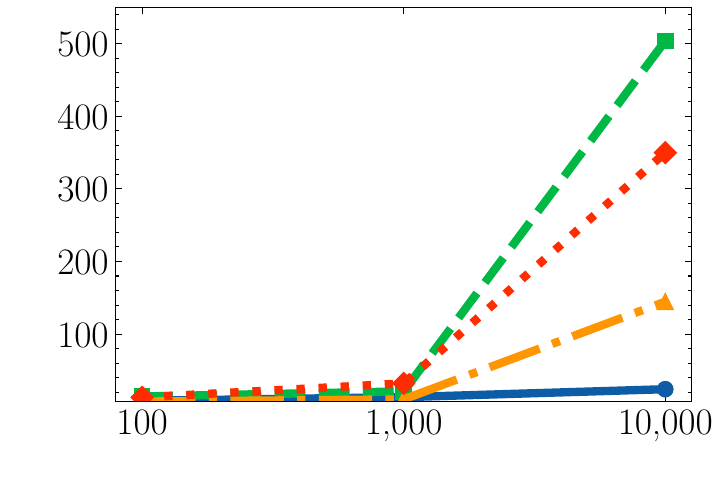}

        \vspace{-1em}
        \caption{\centering Centralized 30$\times$}
        \vspace{0.2em}
    \end{subfigure}
    \hfill
    \begin{subfigure}[t]{0.244\textwidth}
        \centering
        \includegraphics[width=\textwidth]{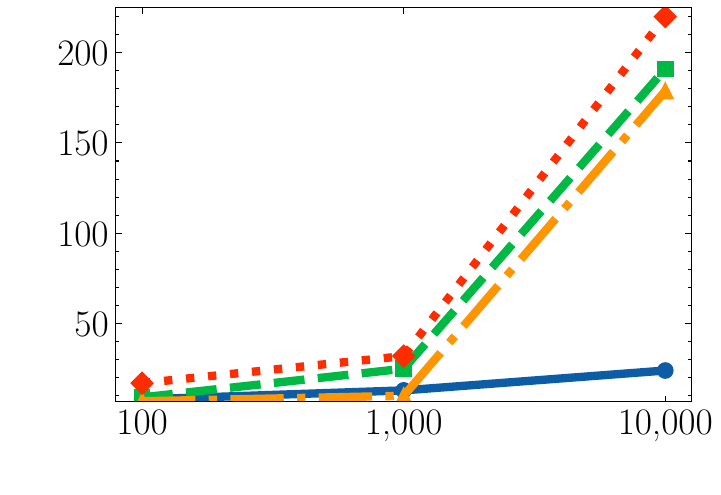}

        \vspace{-1em}
        \caption{\centering Centralized 50$\times$}
        \vspace{0.2em}
    \end{subfigure}
    % \hfill
    \begin{subfigure}[t]{0.244\textwidth}
        \centering
        \includegraphics[width=\textwidth]{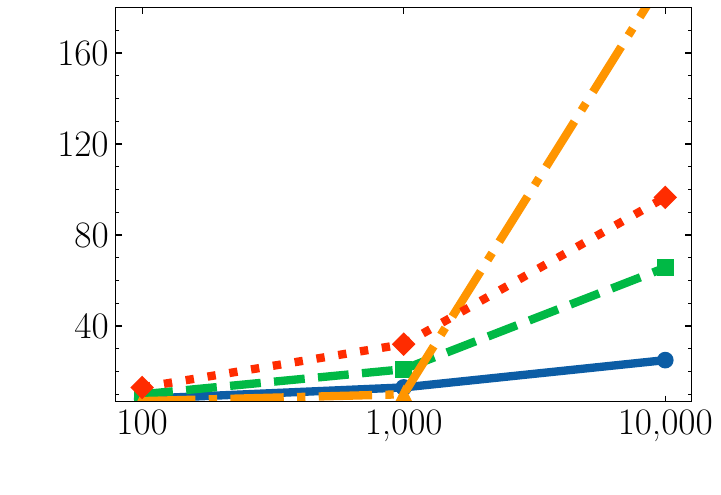}

        \vspace{-1em}
        \caption{\centering Centralized 100$\times$}
        \vspace{0.2em}
    \end{subfigure}
    \begin{subfigure}[t]{0.244\textwidth}
        \centering
        \includegraphics[width=\textwidth]{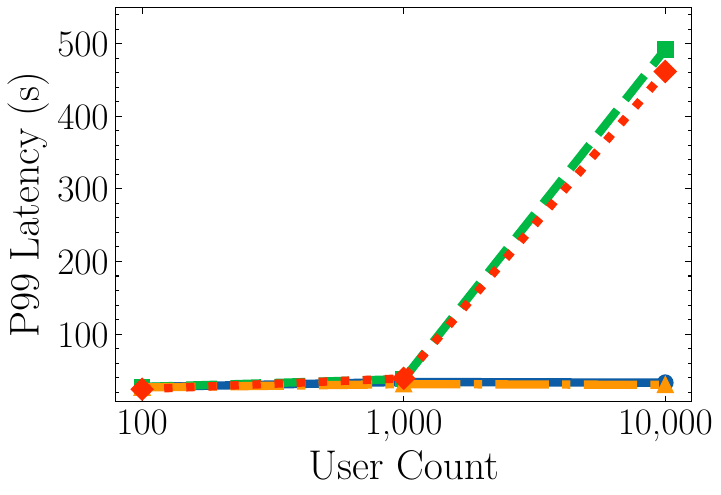}

        \vspace{-0.5em}
        \caption{\centering Distributed 10$\times$}
    \end{subfigure}
    \hfill
    \begin{subfigure}[t]{0.244\textwidth}
        \centering
        \includegraphics[width=\textwidth]{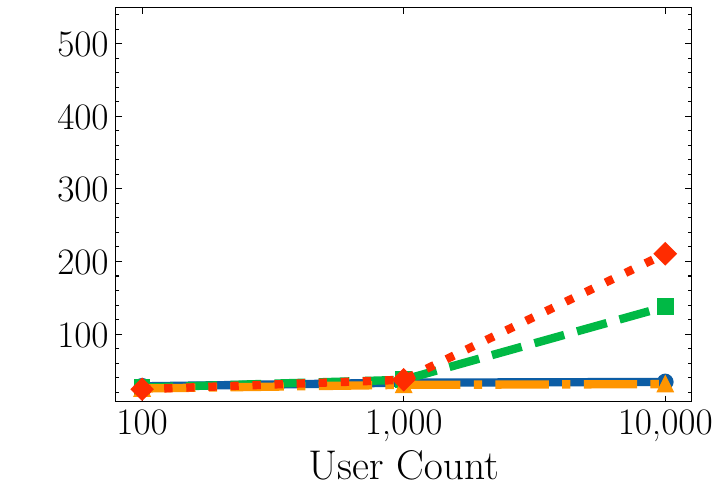}

        \vspace{-0.5em}
        \caption{\centering Distributed 30$\times$}
    \end{subfigure}
    \hfill
    \begin{subfigure}[t]{0.244\textwidth}
        \centering
        \includegraphics[width=\textwidth]{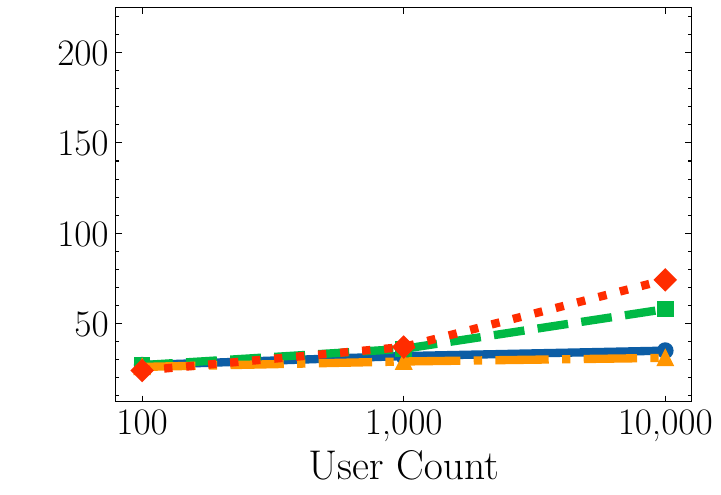}

        \vspace{-0.5em}
        \caption{\centering Distributed 50$\times$}
    \end{subfigure}
    \hfill
    \begin{subfigure}[t]{0.244\textwidth}
        \centering
        \includegraphics[width=\textwidth]{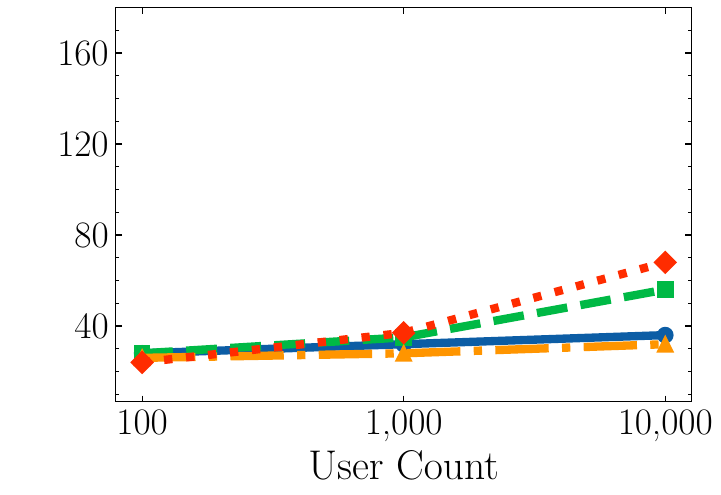}
        
        \vspace{-0.5em}
        \caption{\centering Distributed 100$\times$}
    \end{subfigure}
    
    \caption{P99 request latency estimated over 100,000 simulation steps ($\approx$28\,h) for four inference task types. The x-axis denotes the user count $|\mathcal{I}|$, where each user independently transitions between online and offline states over time. Rows correspond to scheduling policy: \textbf{centralized} (a--d) and \textbf{distributed} (e--h). Columns correspond to server capacity relative to a user node: 10, 30, 50, 100$\times$.
    }
    \label{fig:p99_latency}

\end{figure*}

\begin{figure*} [!ht]
    \centering
    \begin{subfigure}{0.49\textwidth}
        \centering
        \includegraphics[width=\textwidth]{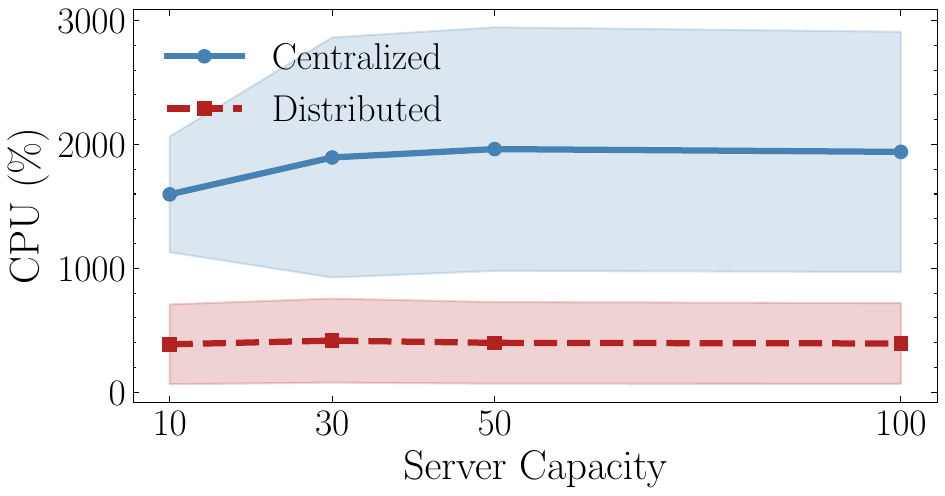}
    \end{subfigure}
    \hfill
    \begin{subfigure}{0.49\textwidth}
        \centering
        \includegraphics[width=\textwidth]{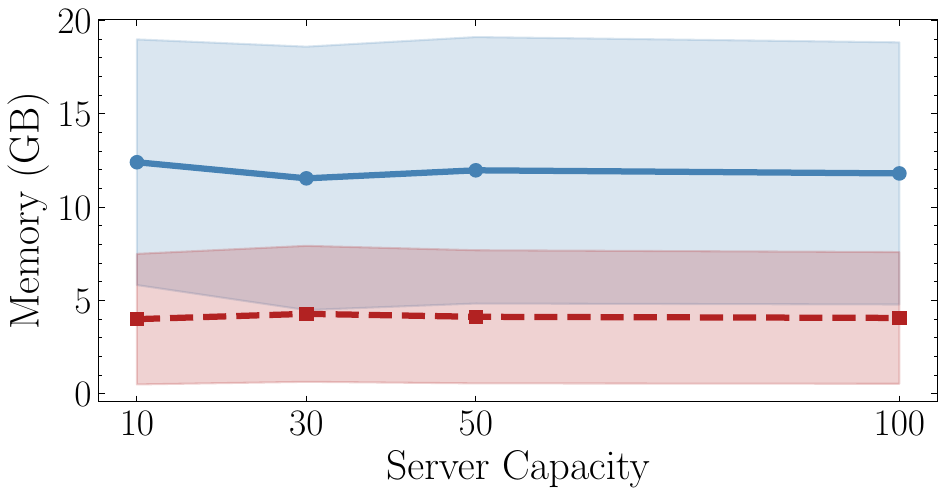}
    \end{subfigure}

    \caption{Server CPU and memory usage estimated over 100,000 simulation steps ($\approx$28\,h) under \textbf{centralized} and \textbf{distributed} scheduling policies, with user count $|\mathcal{I}|$ fixed at 1,000. Curves show the mean, and shaded regions show the standard deviation. The x-axis denotes server capacity relative to a user node.
    }

    \label{fig:resource_usage}
\end{figure*}

\section{Case Study Evaluation}\label{sec:evaluation}

\subsection{Evaluation Setup}

We implemented the model to simulate collaborative distributed inference. The implementation and experimental configuration are available in a GitHub repository.\footnote{\url{https://github.com/AlfredsLapkovskis/CollaborativeInference}} We modeled user availability and request durations using parametric survival distributions. We fitted generative ARMA~\cite{box2015time} models to device measurements to represent allocated resources and resource consumption, after resampling the traces to 1\,s, standardizing and transforming them, and selecting model orders by AIC. We modeled execution-phase durations using discrete-time hazard models~\cite{singer2003applied} fitted to measurements across different resource levels, batch sizes, and subtask types. For simplicity, we assumed i.i.d.\ user behavior and resources. We defined four task types: \emph{light} and \emph{heavy}, each with 5 or 10 subtasks, sampled with probabilities $\{0.4,0.3,0.2,0.1\}$. We evaluated \emph{centralized} and \emph{distributed} scheduler policies across varying server capacities $\mathbf{a}_{\dagger}$ and user counts $|\mathcal{I}|$, measuring P99 latency and server resource consumption over a ${\approx} 28$\,h simulation.
The distributed policy assigns subtasks to user nodes randomly, subject to simple feasibility heuristics.

\subsection{Latency Analysis}

Fig.~\ref{fig:p99_latency} shows P99 latencies estimated from simulation. As the number of users grows, the distributed policy progressively outperforms the centralized one: beyond a certain user count, the server becomes a bottleneck under centralized scheduling, and even naive subtask distribution across user nodes yields lower latency. This demonstrates the practical feasibility of distributing inference subtasks to user devices.

The distributed policy serves as an informative lower bound on achievable performance.
The results suggest that an effective policy would dynamically shift computation toward user devices as load increases, concentrating it on the server under lighter demand. Designing and evaluating such policies, however, requires a formal system model. This is precisely where our contribution lies. Our proposed framework enables simulation under varying conditions, policy evaluation, and optimization (cf.\ Definition~\ref{def:scheduler_policy_optimization}) via RL methods, since the factorized Markov structure is directly compatible with RL. Its modular nature further allows extending the model to new resource types, task structures, or node behaviors without redesigning the framework.

\subsection{Resource Efficiency}

Fig.~\ref{fig:resource_usage} shows dedicated server resource consumption with 1,000 users. The distributed policy consistently achieves lower CPU and memory usage across all server capacity levels. Notably, at 1,000 users, both policies yield comparable P99 latency (Fig.~\ref{fig:p99_latency}), yet the distributed policy incurs considerably lower server resource costs. Offloading subtasks to user devices thus provides efficiency gains beyond latency alone, motivating Definition~\ref{def:dedicated_resource_optimization}: given a fixed policy, one can minimize nominal server resources subject to QoS constraints, directly reducing infrastructure costs.

\section{Discussion} \label{sec:discussion}

Our case study demonstrates that the proposed framework can represent and simulate a sophisticated distributed computing system of practical relevance. The results show that centralized scheduling becomes a bottleneck at scale, whereas even random distribution of computation across user devices can substantially reduce request latency and dedicated server resource consumption. The gains achieved by this simple policy suggest that more carefully designed scheduling policies could yield further improvements. 

Evaluating and optimizing such policies without a system model, however, is challenging due to the system's stochasticity, dynamism, heterogeneity, and scale. Model-free approaches would require either extensive experience collected from the deployed system (e.g., for off-policy RL) or online interaction to train an adaptive policy. Both alternatives require substantial interaction with the real system. This makes policy development time-consuming, resource-intensive, and potentially disruptive to the provided service. 

A system model can reduce this dependence on real-world interaction by enabling policies to be evaluated and optimized through simulation. This is where our framework contributes: it provides a uniform probabilistic representation of heterogeneous system components while allowing each component's dynamics to be modeled independently using an appropriate modeling technique. Its sparse factorization keeps the resulting high-dimensional model tractable. The framework also connects distributed computing system modeling with established methods from Markov chain theory and RL.

\subsection{Current Limitations and Future Work} \label{sec:limitations}

The present work has several limitations that we aim to address in future work. First, we demonstrate the framework through a single case study of collaborative distributed AI inference. Although this case study represents a timely and practically relevant application, additional case studies are needed to assess the framework's applicability to other distributed computing systems. Second, our evaluation so far considers only two manually designed scheduling policies. While these policies already reveal meaningful performance trade-offs and demonstrate the framework's usefulness for simulation and policy evaluation, they are not intended to be optimal. We formulate scheduler policy and dedicated resource optimization problems and show how they can be expressed using the proposed model, but solving them requires substantial further investigation and is therefore within the scope of our future work. 

The case study implementation also currently relies on several simplifying assumptions. In particular, we assume homogeneous user devices to avoid introducing additional mechanisms for handling fairness among participating users. These simplifications were necessary to keep the implementation manageable, as the instantiated model already captures a broad range of interacting system components and dynamics. Future work will relax these assumptions by considering heterogeneous users and fairness-aware decision-making. We will also investigate adaptive scheduling and dedicated resource optimization, apply the framework to additional case studies, and validate the instantiated models and their simulation outcomes against observations from real-world systems.

\section{Conclusion} \label{sec:conclusion}

This paper introduced a general framework for modeling distributed computing systems as structured generative Markov models. The framework is particularly relevant to the heterogeneous, stochastic, and dynamic computing continuum systems. Its two-level sparse factorization provides a uniform probabilistic representation of diverse system components while allowing their dynamics to be modeled independently using appropriate techniques. This modular structure also makes the framework readily extensible, as new system components, entity types, and dependencies can be incorporated without redesigning the overall model. We demonstrated the framework through a comprehensive model of collaborative distributed AI inference. The results show that centralized scheduling becomes a bottleneck at scale, whereas distributing computation across user devices can reduce latency and dedicated server resource consumption. More broadly, the case study demonstrates the framework's utility for simulation, policy evaluation, and optimization. Future work will apply the framework to additional distributed computing systems, extend the case study to heterogeneous users and fairness-aware decision making, investigate adaptive scheduling and dedicated resource optimization, and validate the instantiated models against observations from real systems.

\section*{Acknowledgement}

The computations were enabled by resources provided by the National Academic Infrastructure for Supercomputing in Sweden (NAISS), partially funded by the Swedish Research Council through grant agreement no. 2022-06725.

\bibliographystyle{IEEEtran}
\bibliography{ref.bib}

\end{document}